\documentclass[pdflatex,sn-mathphys-num]{sn-jnl}

\usepackage{graphicx}%
\usepackage{multirow}%
\usepackage{amsmath,amssymb,amsfonts}%
\usepackage{amsthm}%
\usepackage{mathrsfs}%
\usepackage[title]{appendix}%
\usepackage{xcolor}%
\usepackage{textcomp}%
\usepackage{manyfoot}%
\usepackage{booktabs}%
\usepackage{algorithm}%
\usepackage{algorithmicx}%
\usepackage{algpseudocode}%
\usepackage{listings}%
\usepackage{tablefootnote}
\usepackage{float}

\theoremstyle{thmstyleone}%
\theoremstyle{thmstyletwo}%

\theoremstyle{thmstylethree}%

\begin{document}

\title[Assessing Computer Science Student Attitudes]{Assessing Computer Science Student Attitudes Towards AI Ethics and Policy}


\author[1]{\fnm{James} \sur{Weichert}}
\author[2]{\fnm{Dayoung} \sur{Kim}}
\author[2]{\fnm{Qin} \sur{Zhu}}
\author[3]{\fnm{Junghwan} \sur{Kim}}
\author*[1]{\fnm{Hoda} \sur{Eldardiry}}\email{hdardiry@vt.edu}

\affil[1]{\orgdiv{Department of Computer Science}, \orgname{Virginia Tech}, \orgaddress{\city{Blacksburg}, \state{VA}}}
\affil[2]{\orgdiv{Department of Engineering Education}, \orgname{Virginia Tech},
\orgaddress{\city{Blacksburg}, \state{VA}}}
\affil[3]{\orgdiv{Department of Geography}, \orgname{Virginia Tech},
\orgaddress{\city{Blacksburg}, \state{VA}}}


\abstract{As artificial intelligence (AI) grows in popularity and importance—both as a domain within broader computing research and in society at large—increasing focus will need to be paid to the ethical governance of this emerging technology. The attitudes and competencies with respect to AI ethics and policy among post-secondary students studying computer science (CS) are of particular interest, as many of these students will go on to play key roles in the development and deployment of future AI innovations. Despite this population of computer scientists being at the forefront of learning about and using AI tools, their attitudes towards AI remain understudied in the literature. In an effort to begin to close this gap, in fall 2024 we fielded a survey ($n=117$) to undergraduate and graduate students enrolled in CS courses at a large public university in the United States to assess their attitudes towards the nascent fields of AI ethics and policy. Additionally, we conducted one-on-one follow-up interviews with 13 students to elicit more in-depth responses on topics such as the use of AI tools in the classroom, ethical impacts of AI, and government regulation of AI. In this paper, we describe the findings of both the survey and interviews, drawing parallels and contrasts to broader public opinion polling in the United States. We conclude by evaluating the implications of CS student attitudes on the future of AI education and governance.

}




\maketitle

\section{Introduction}\label{sec1}

The rapid development and proliferation of artificial intelligence (AI) technologies—from personal assistants to large language models to self-driving cars—is evident across research publications, patents, and public and private investment \cite{maslej_ai_2024}. As a result of the widespread availability of AI-enabled consumer products, Americans—especially young Americans between 18 and 29 years old—are becoming more familiar with `AI', both as an abstract concept and as manifested in specific tools like ChatGPT or Siri \cite{mcclain_americans_2024, jackson_googleipsos_2025}. At the same time, governments and companies alike are beginning to develop roadmaps or implement policies to govern the use of AI \cite{noauthor_exec_2023, weichert_perceptions_2025, jobin_global_2019, auld_governing_2022}. 

Similarly, AI is increasingly featuring as a key topic in many computing ethics courses \cite{weichert_i_2025, weichert_evolution_2025}. The emergence of \textit{AI ethics} in the classroom comes amid a growing academic consensus around the need to emphasize the responsible use of AI technologies through ethics education \cite{borenstein_emerging_2021, lauer_you_2021, anderson_ai_2021, floridi_ethics_2023}. Moreover, nascent \textit{AI policy} efforts suggest that technical AI practitioners will in future be expected to operationalize abstract ethical principles and guidelines into the code underlying AI models \cite{kim_exploring_2023, kim_toward_2024}. This expectation is made clear by former U.S. President Biden's 2023 executive order on AI, which called for an ``AI talent surge,'' and the training of the federal workforce on ``AI issues,'' \cite{noauthor_exec_2023}.

Yet for the most part, discussions about how to oversee and regulate AI are absent from the computer science (CS) classroom \cite{weichert_evolution_2025}. More fundamentally, the attitudes of the CS students in these classes are understudied, even though young people—and young computer scientists in particular—are more likely to regularly use AI tools in their work and daily lives. We posit that these students inhabit a unique position in the AI landscape. On the one hand, they are \textit{AI learners} in machine learning, AI, and computing ethics courses, studying the underlying technical mechanics driving state-of-the-art AI models. Simultaneously, they are \textit{AI consumers}, regularly using the AI technologies they learn about in class and developing new workflows to integrate AI tools into their lives.

These `dual roles' present both opportunities—as students can easily apply classroom content to their use of AI, or vice versa—and challenges—as students' ethical attitudes towards AI may be formed in large part through the unguided use of AI tools. For this reason, understanding how CS students use and view AI technologies is crucial for identifying underdeveloped competencies relating to AI ethics and policy, adapting to new AI workflows, and revising existing ethics curricula to better promote the responsible development and use of AI. However, we find that there currently exists no comprehensive assessment of usage, attitudes, and competencies related to AI specifically among the CS student population. 

To address this gap in the literature, in fall 2024 we administered an online survey ($n = 117$) to undergraduate and graduate students in a variety of CS courses at a large public university in the southeastern United States. The survey assesses student attitudes towards AI in general, the ethical impacts of AI, and AI policy. We complement these quantitative findings with semi-structured follow-up interviews ($n=13$), which provide deeper insights into how and why students use AI tools, as well as student interest in AI policy careers. In doing so, we aim to address three research questions related to how CS students view and interact with emerging AI technologies:

\begin{itemize}
    \item \textbf{RQ1} How and how often do CS students use AI tools in their university studies and in day-to-day life?
    \item \textbf{RQ2} What are CS students' attitudes towards emerging AI technologies, both in general and with respect to AI's ethical impacts?
    \item \textbf{RQ3} To what extent are CS students interested in and prepared for a job involving AI ethics, regulation, policy compliance, and/or policymaking?
\end{itemize}

By synthesizing our quantitative and qualitative data sources, we identify common themes summarizing students' attitudes with respect to AI. Although our findings are most relevant in the context of AI education and CS curricula, we also anticipate that a better understanding of how young computer scientists think about AI will be insightful for both industry and government leaders.

\section{Literature Review}

\subsection{AI Ethics and Policy}

The growing interest in AI in both academic and professional contexts has simultaneously drawn attention to the ways in which the challenges and social impacts of AI can be conceptualized and addressed. This focus has resulted in the emergence of \textit{AI ethics} as a specific and distinct subfield within computing ethics \cite{lauer_you_2021, borenstein_emerging_2021, anderson_ai_2021, hauer_importance_2022, floridi_ethics_2023}. Borrowing from a definition of ``machine ethics'' from Anderson and Anderson \cite{anderson_machine_2007}, we define \textit{AI ethics} as the study and practice of aligning the behavior of AI systems with the norms and outcomes desired by humans. Recognizing that there is still much disagreement about what these norms are and how they should apply to AI—evidenced by the multitude of different AI ethics principles identified by Jobin et al. \cite{jobin_global_2019} and skepticism about the efficacy of ``AI ethics'' in the first place \cite{munn_uselessness_2023}—we focus specifically on how AI stakeholders think about and seek to address the impacts of AI on society. Whereas the ethics \textit{of} AI concerns itself with which ethical principles should govern AI development and how principles can be translated to practice \cite{schiff_principles_2020, kim_exploring_2023}, \textit{AI ethics} seeks to provide a framework for thinking about AI harms and interventions. The latter can therefore better adapt to changing technological capabilities, AI use cases, and societal expectations regarding AI use.

Whereas both ``ethics of AI'' and ``AI ethics'' appear frequently throughout AI literature, the term ``AI policy'' is much less prevalent. We view \textit{AI ethics} as being naturally complemented by \textit{AI policy}. Where the former is primarily concerned with conceptualizing the challenges and impacts of AI, the latter is concerned with how to effectively operationalize normative ethical priorities through regulation, governance, and technical tools \cite{kim_exploring_2023, kim_toward_2024, morley_what_2020}. Although the term ``policy'' often suggests government regulation, this is not always the case. Private companies may choose to implement policies governing their AI systems for a variety of reasons, including adapting to consumer opinion or preempting stricter AI controls by the government \cite{auld_governing_2022, floridi_translating_2019}. As such, \textit{AI policy} is best described as the study of the structures and processes through which policies and regulations governing AI are developed, implemented, and enforced. Moreover, in democratic societies in general, understanding the public's multifaceted perceptions of emerging technologies—and how those perceptions may influence the policymaking process \cite{kim_examination_2021, kim_role_2024}—is key to formulating effective technology policies.

Despite the emergence of AI ethics as a field of inquiry that has generated increasing scholarly attention and recent government regulatory developments, the adoption of AI ethics content in undergraduate CS programs is not widespread \cite{weichert_i_2025}, and very few courses incorporate discussions about the AI policy and regulatory landscape into their curriculum \cite{weichert_evolution_2025}.

\subsection{Use of AI Tools in CS Education}

The expanded availability of commercial AI tools—like OpenAI's ChatGPT large language model (LLM)—has precipitated the integration of AI into workflows to accomplish a wide range of tasks. While a 2024 Pew Research Center \textit{American Trends Panel} survey \cite{mcclain_americans_2024} found that, overall, only 23\% of surveyed Americans reported having used ChatGPT, this figure was almost double (43\%) among respondents aged 18-29, having increased from 33\% in 2023. The familiarity of post-secondary students with AI technology has prompted educators, particularly in computer science and data science, to explore how to integrate AI tools into their courses, encouraging cautious experimentation \cite{chan_comprehensive_2023}. Recent papers have evaluated the benefits and challenges of using AI as a pedagogical tool in computing classrooms \cite{qureshi_chatgpt_2023, shen_implications_2024}, for example through the creation of custom AI-enabled course infrastructure for large-enrollment introductory programming courses \cite{liu_teaching_2024, zamfirescu-pereira_61a_2024}. These studies tend to conclude that the use of AI tools in CS teaching is generally effective and resource efficient, but also that appropriate `guardrails' are needed to promote the responsible use of AI in coursework \cite{liu_teaching_2024}. 

Less is known about how students approach the use of AI tools in higher education when no explicit encouragement or guidance is provided by instructors, or where the use of AI on assignments constitutes a `grey area' with respect to academic honesty. A survey of students and faculty on generative AI use fielded in fall 2023 by Kim et al. \cite{kim_examining_2025} found that only 28\% of student respondents at a large public university in the U.S. had never used generative AI tools (e.g. ChatGPT), while regular usage was generally higher for STEM students (vs. non-STEM students) and for male students (vs. female students). Ngo \cite{ngo_perception_2023} and Singh et al. \cite{singh_exploring_2023} both conduct surveys assessing the attitudes of university students (in Vietnam and the U.K., respectively) towards ChatGPT, finding that students are generally aware of the limitations of ChatGPT and many have concerns about misusing the LLM in their academic activities. Ngo also conducts semi-structured interviews with a subset of students, identifying common ChatGPT uses, including searching for information, brainstorming ideas, and providing feedback on student writing \cite{ngo_perception_2023}. The potential use of AI for generating academic writing has generated concern among educators around an anticipated rise in academic dishonesty \cite{cotton_chatting_2024}, especially as AI-generated content detection schemes have yet to live up to expectations \cite{chaka_detecting_2023, weichert_dupe_2024}. However, a study of high schooler ChatGPT use by Levine et al. \cite{levine_how_2024} found that a large majority of ChatGPT uses in the context of writing involved either structuring students' own ideas or reviewing and providing feedback on student writing. The authors therefore conclude that AI tools can be used ``productively and responsibly'' in supporting student learning, even in the context of academic writing assignments. 

Given that much is still unknown about how students, especially in computer science departments, organically interact with available AI technologies, a key aim of the present study is to highlight common AI tools and use cases identified through both survey data and semi-structured follow-up interviews. This context, in turn, is valuable for deciphering how CS students form or adapt their attitudes towards AI more generally through their interactions with AI systems in classroom and non-classroom settings.

\subsection{Attitudes Towards AI}

Numerous studies in recent years have proposed and fielded surveys to measure respondents' attitudes towards AI across a variety of scenarios and ethical dimensions \cite{zhang_us_2020, merenkov_public_2021, liehner_perceptions_2023, rojahn_american_2023, maslej_ai_2024}. Public opinion firms Pew and Ipsos both regularly field questions about AI in their surveys \cite{mcclain_americans_2024}, with the 2025 Google/Ipsos \textit{Multi-Country AI Survey} \cite{jackson_googleipsos_2025} finding that only 15\% of respondents across 21 countries didn't know ``anything at all'' about ``artificial intelligence (AI) tools and applications'' while the plurality of respondents reported being ``Mostly excited about the possibilities, but somewhat concerned about the risks'' when thinking about AI. Among academic surveys, the \textit{General Attitudes towards Artificial Intelligence Scale} (GAAIS) by Schepman and Rodway \cite{schepman_general_2023} stands out as being especially robust, with Likert-scale rating questions featuring both positive (e.g. ``AI is exciting'') and negative (e.g. ``I think artificially intelligent systems make many errors'') statements about AI. For our purposes, however, the GAAIS is too narrowly focused on ethical impacts of AI, does not assess \textit{self-efficacy} related to AI ethics and policy, and contains significantly more questions than we would like for our lightweight survey design.   

Of particular relevance to our research are those studies which develop surveys of attitudes towards AI designed specifically for students \cite{ghotbi_moral_2021, jang_development_2022, hooper_analysis_2022, suh_development_2022, chan_students_2023, jurado_students_2023, strzelecki_students_2024, asio_predictors_2024, kharroubi_knowledge_2024, kim_examining_2025}. Among these, Jang et al. \cite{jang_development_2022} study student attitudes across five ethical principles (fairness, transparency, nonmaleficence, privacy, and responsibility), finding key differences in responses between male and female students as well as between students with prior AI education and students without. Similarly, Asio and Gadia \cite{asio_predictors_2024} report a correlation between students' literacy and self-efficacy with respect to AI and their attitudes towards AI. These findings in particular suggest that the attitudes of CS students as a standalone group should be investigated, as these students are the most likely to be familiar with AI technologies on a technical level and to use AI in daily tasks. Ghotbi and Ho \cite{ghotbi_moral_2021} ask students to rank ethical impacts of AI according to which they perceive to be the most important. The authors find that by far the most selected response was ``increasing unemployment'' due to AI-driven automation. Finally, the survey of U.S. college students by Kim et al. \cite{kim_examining_2025} serves as an important baseline for our study, as it includes questions related to AI use. Notwithstanding the contributions of these papers, however, we find that we must proceed with our own survey instrument in order to investigate the particular aspects of AI attitudes that most interest us. To our knowledge, our study (including our initial pilot in \cite{weichert_computer_2024}) is the first study of college-level CS students focused on AI ethics and policymaking \textit{self-efficacy} and \textit{attitudes} towards AI regulation.

Specifically, our survey instrument covers attitudes and competencies across three sections: \textit{General Attitudes Towards AI}, \textit{AI Ethics}, and \textit{AI Regulation and Policy}. Our experiences in piloting this survey among students in an upper-level undergraduate CS class \cite{weichert_computer_2024} give us confidence in the thematic coverage of our instrument. In contrast to the pilot, however, we extend our survey to include two questions modeled on existing studies. First, the \textit{General Attitudes Towards AI} section of the survey includes the Likert statement ``Much of society will benefit from a future full of AI,'' which is based on Schepman and Rodway's \cite{schepman_general_2023} GAAIS. Utilizing this question will allow us to compare the response distribution between our sample of CS students and the sample of the general UK population from Schepman and Rodway. Second, we adapt from Ghotbi and Ho \cite{ghotbi_moral_2021} a question asking respondents to select from among potential ethical issues related to AI the one which they believe will be ``most important'' in the future. We include this question twice in our survey, asking about student perceptions of relevant ethical issues \textit{now} and in the \textit{future}. We expand upon the enumerated ethical issues in Ghotbi and Ho by adding \textit{autonomous decision-making}, \textit{hallucination and/or false information}, and \textit{deepfake (synthetic) content} as additional options.

\section{Methodology}

\subsection{Study Population}

The study population for this Institutional Review Board-approved research consists of undergraduate and graduate students enrolled in computer science (CS) courses at a large public research university in the southeastern United States. We choose to narrow our sample to specifically those students engaging with curriculum on AI, machine learning, and/or computing ethics. After compiling a list of all courses being offered in fall 2024 relating to AI, ML or CS ethics, emails were sent to each instructor asking to advertise the survey through an email to the class, an in-person class announcement, or an online announcement on the course's learning management system (LMS). 

80 of the 117 survey respondents (68\%) were undergraduate students, while 26\% were master's students and 6\% were PhD students. These proportions represent a slight oversampling of undergraduate students compared to the composition of undergraduate students in the CS department (64\%) and an undersampling of PhD students (13\%). However, we maintain that our survey sample represents a robust cross-section of student academic levels and courses.

\subsection{Survey Instrument}

Our survey consists of three sections:\textit{General Attitudes Towards AI}, \textit{AI Ethics}, and \textit{AI Regulation and Policy}. The full instrument is listed in Appendix \ref{secA1}. The survey's design is strongly influenced by existing survey instruments from Kim et al. \cite{kim_examining_2025}, Jang et al. \cite{jang_development_2022}, Ghotbi et al. \cite{ghotbi_moral_2021}, Hooper and Fletcher \cite{hooper_analysis_2022}, and Shepman and Rodway \cite{schepman_general_2023}. Each of the survey's three sections consists of eight Likert-scale questions on a 1 to 5 scale. Additionally, the \textit{General Attitudes Towards AI} section includes questions about how often respondents use AI tools and which tools they use, while the \textit{AI Ethics} section asks respondents to select, from a pre-populated list, up to two potential ethical impacts of AI technologies that they worry about. To analyze the quantitative survey data, we visualize the rating distribution of each Likert question (see Appendix \ref{secA3}), computing descriptive statistics to summarize key findings.

\subsection{Follow-Up Interviews}

While completing the survey, respondents were asked if they would like to be contacted via email for a follow-up interview in the coming weeks. As a result of this outreach, we conducted six interviews with undergraduate students and seven interviews with graduate students. These semi-structured follow-up interviews were designed to expand on the topics addressed in the survey, using open-ended questions to guide interviewees without restricting their responses. Each interview lasted approximately 15 minutes and was conducted online via Zoom videoconferencing. The interview focused on our key research strands, including general attitudes towards AI, the use of AI tools, the incorporation of AI ethics in CS education, and interviewees' familiarity with and attitudes towards AI regulation. These topics were split over six broad interview questions (IQs), listed in Appendix \ref{secA2}. After the interviewee responded to the initial question, the interviewer asked shorter and more targeted follow-up questions to clarify aspects of the student's response or elicit specific examples. Each interview was audio recorded and automatically transcribed for use during our analysis.

\section{Findings}

\subsection{Student Attitudes Survey}

Below, we summarize key findings from our survey of 117 undergraduate and graduate CS students. Response distributions for the Likert-scale questions in each survey section are visualized in Figures \ref{fig:general-likert}, \ref{fig:ethics-likert}, \ref{fig:policy-likert} in Appendix \ref{secA3}, respectively.

\subsubsection{General Attitudes}

\begin{figure}
    \centering
    \includegraphics[width=1\linewidth]{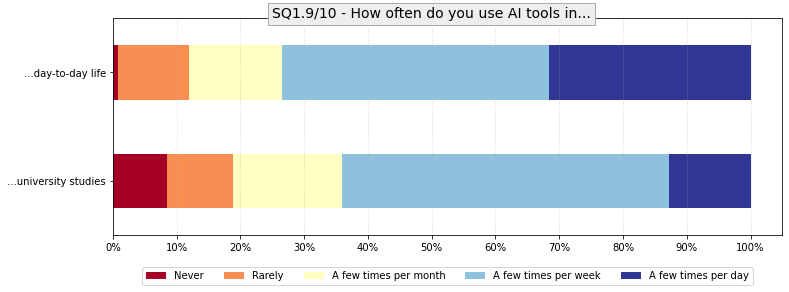}
    \caption{Frequency of CS student AI tool use in day-to-day life and university studies.}
    \label{fig:frequency}
\end{figure}

Overall, student perceptions of their understanding of AI, the benefits of AI to themselves and society, and the effect of using AI on their productivity are very positive, with over 60\% of respondents agreeing or strongly agreeing with all eight Likert questions in this section (Figure \ref{fig:general-likert}). Agreement was particularly high for SQ1.1, which assesses students' familiarity with AI (86\% agree or strongly agree), and for SQ1.8, with 80\% agreeing or strongly agreeing that ``using AI tools increases my productivity.'' This high level of agreement is unsurprising in the context of how often students report using AI tools. Figure \ref{fig:frequency} shows strong use frequency in the context of university studies (SQ 1.10), with 51\% of students using AI at least a few times per week and 13\% at least a few times per day. AI usage in day-to-day life (SQ 1.9) is even higher, with nearly a third (32\%) of respondents using AI a few times per day and 42\% at least a few times per week. The high rate of adoption of AI in relation to university coursework is perhaps surprising in comparison with the Kim et al. survey \cite{kim_examining_2025} of generative AI use in the overall undergraduate population, which found that less than 30\% of students used generative AI at least multiple times per week in academic settings. While Kim et al. found slightly higher usage among STEM students, the 64\% of CS students we find that use AI at least a few times per week suggests that the CS vs. non-CS student divide may be a more significant distinction in contrasting AI adoption than STEM vs. non-STEM. We discuss the implications of this high usage, incorporating additional context from our follow-up interviews, in Section \ref{discussion}.

\begin{figure}
    \centering
    \includegraphics[width=0.8\linewidth]{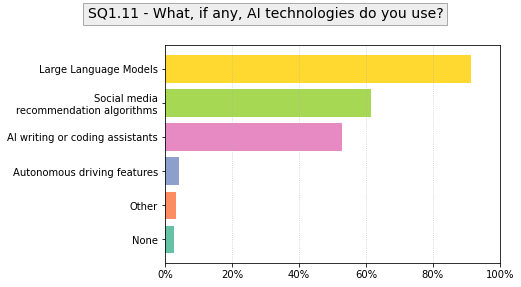}
    \caption{Percentage of survey respondents who indicated using specific AI technologies.}
    \label{fig:ai-tech}
\end{figure}

Survey question 1.11 also provides an overview of what AI tools are being used by students. Among the five explicit AI categories (\textit{No AI}, \textit{Large Language Models (LLMs)}, \textit{AI writing or coding assistants}, \textit{Autonomous driving features}, and \textit{Social media recommendation algorithms}), LLMs, social media recommenders, and AI assistants were each selected by over 50\% of respondents. As shown in Figure \ref{fig:ai-tech}, few students indicated use of self-driving features, and even fewer indicated they use no AI tools at all, confirming the high level of AI adoption evidenced by SQ1.9 and SQ1.10. The frequent use of both types of generative AI tools provides further evidence for the hypothesis from our pilot study \cite{weichert_computer_2024} that LLMs and embedded AI assistants fulfill distinct use cases in an `AI workflow.' We expanded upon this thread in our follow-up interviews (see Section \ref{interviews}), which provided an indication for \textit{how} students use these AI tools. We discuss further implications of a high level of AI tool adoption for CS pedagogy in Section \ref{discussion}.

Finally, the \textit{General Attitudes Towards AI} section also included a question adapted from Schepman and Rodway's \cite{schepman_general_2023} GAAIS, asking respondents to rate their agreement with the statement ``Much of society will benefit from a future full of AI,'' Using the same 5-point Likert scale as Schepman and Rodway, we can directly compare the response distributions between the CS student and general populations. Whereas Shepman and Rodway find a mean response for this question of 3.438 out of 5, our SQ1.5 returned a higher response mean of 3.641. Given that the response ratings are not normally distributed, we proceed with a one-sample Wilcoxon signed-rank test (a non-parametric generalization of the t-test for non-normal distributions \cite{armitage_nonparametric_2005}), comparing our response distribution to $\mu = 3.438$. With a p-value less than 0.001 we can, with high confidence, reject the null hypothesis and conclude that the difference between Schepman and Rodway's sample of the general population (in the U.K.) and our sample of CS students is statistically significant. While not identical, GAAIS also asks respondents to rate interest ``in using artificially intelligent systems in my daily life,'' The mean for this statement, 3.418, is lower than the mean of 3.897 for SQ1.6 (``AI tools are helpful to me in my day-to-day life''). We conduct a Wilcoxon signed-rank test as above and likewise arrive at a p-value less than 0.001.

\subsubsection{AI Ethics}

The response distribution for Likert questions related to AI ethics (Figure \ref{fig:ethics-likert}) is overall less uniform than for general questions about AI, with respondent opinion more evenly divided on questions about the ethics of existing AI technologies. While only 17\% of respondents disagree or strongly disagree with the statement ``In general I think existing AI tools are ethical'' (SQ 2.1)—compared to 40\% agree or strongly agree—almost as many respondents disagree with the statement ``I believe that most developers of AI tools design their systems with ethics in mind'' (SQ 2.2) as agree—36\% vs. 40\%. Despite somewhat split opinion of the `ethics' of AI technologies, respondents were mostly in agreement in their concern about the ethical \textit{impacts} of AI technologies. Nearly 70\% of respondents agreed or strongly agreed that they ``worry about the ethical impact of current AI technology,'' (SQ 2.3). This agreement increases to over 75\% with respect to the potential impacts of \textit{future} AI technologies (SQ 2.4), with the percentage of respondents strongly agreeing doubling to 31\%.

\begin{figure}
    \centering
    \includegraphics[width=1\linewidth]{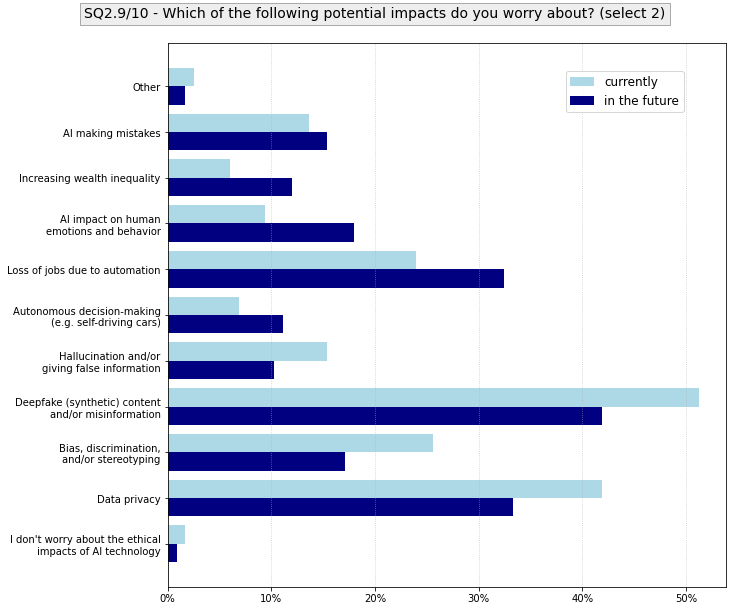}
    \caption{Percentage of survey respondents who indicated concern about specific impacts of AI now and in the future.}
    \label{fig:worries}
\end{figure}

When asked in SQ2.9 and 2.10 about potential impacts of AI that concerned respondents (Figure \ref{fig:worries}), \textit{deepfake content and/or misinformation} and \textit{data privacy} were the most commonly selected responses. Interestingly, both of these impacts see higher concern when talking about \textit{current} AI impacts compared to the impacts respondents think will be important in the \textit{future}. One possible explanation for this phenomenon is that students assume that these impacts are being or will shortly be addressed through technical improvements (e.g. deepfake content detectors or widespread adoption of techniques like differential privacy). As such, there may be an anticipation that these concerns will be mitigated (or resolved entirely) in the future. By contrast, concern about \textit{AI impact on human emotions and behavior} and \textit{loss of jobs due to automation} increases in the future context, suggesting that students are more distantly concerned about the broad-scale impacts of pervasive AI tools on society and human interaction.

With respect to their own ethical competencies, students are confident in their ability to explain to someone unfamiliar with AI ``how AI can make biased or harmful decisions'' (nearly 90\% agree or strongly agree with SQ 2.5). Students likewise recognize how their AI or ML courses have integrated discussions about the ethics of AI into the curriculum, and over 60\% of respondents find that ``sufficient attention'' is paid to AI ethics in their CS education (SQ 2.7). However, there is some disagreement (10\%) and neutral opinions (24\%) among respondents regarding whether their university courses have ``prepared [them] to discuss and mitigate potential harms that AI can cause'' (SQ 2.8). 

\subsubsection{AI Policy and Regulation}

Finally, students were asked to rate their agreement with statements relating to AI policy and the regulation of AI technology (Figure \ref{fig:policy-likert}). A majority of respondents disagreed or strongly disagreed that ``AI technologies are currently adequately regulated by the government,'' (SQ 3.1) while a plurality (47\%) disagreed with the statement ``The U.S. government is doing a good job of balancing promoting innovation and protecting users with respect to AI technology,'' (SQ 3.2). Responses to questions 3.3 and 3.4 suggest that students are concerned both about protecting users and protecting society at large through the regulation of AI technology.

Although we find that respondents have strong opinions regarding AI regulation, only around 40\% of respondents indicated in SQ 3.5 that they tend to follow news about government regulation of AI or technology at large. Even fewer (32\%) expressed interest in AI policy as a potential career path (SQ 3.6). This figure is significantly lower than the 44\% of respondents in our previous study \cite{weichert_computer_2024}, but the inclusion of graduate students—who in large part are beginning to specialize in particular computing domains—in our current study helps to explain this decrease. Nonetheless, our survey results indicate that nearly one third of CS students would be interested in a career related to AI or technology policy, while another quarter could—in our opinion—be persuaded through increased exposure to available career opportunities in technology policy. To effect an increase in student interest in policy-related careers, however, discussions of AI policy and regulation must be further integrated into the computing curriculum. Less than half of respondents indicated that their AI or ML courses included discussions about ``the regulation of AI by the government'' (SQ 3.7), while only 46\% think their courses ``are adequately preparing [them] to engage in conversations about AI policy and regulation,'' (SQ 3.8). We discuss the implications of these findings further in Section \ref{discussion}.

\subsection{Student Interviews}\label{interviews}

Below, we summarize common attitudes and recurring themes that emerged from our semi-structured follow-up interviews with 13 undergraduate and graduate CS students. Each interview participant is referred to using a pseudonymized alias.

\subsubsection{General Views on AI}

Overall, the students we interviewed had exceedingly positive views towards AI in general. When asked whether they considered AI ``currently a net positive or net negative to society,'' 11 of 13 interviewees (85\%) responded with ``net positive'' while only 2 expressed that their views on AI were ``mixed''. However, further discussion of each student's views on AI revealed more nuanced attitudes towards this emerging technology. For example, Nick noted that AI ``has very good uses...but it can also be harmful, especially when people rely on it for everything or when they don't double check the information they get from AI,'' Likewise, Kenny noted the potential for generative AI technologies to stifle human creativity. Others argued that public perception of the capabilities of AI were exaggerated. Spencer thought that ``people are overestimating the variety of the cases that [AI is] useful for...we have not achieved general intelligence yet,'' while Brett viewed the ``tech sector'' as being ``a little bit optimistic on how beneficial [AI] is.''

\subsubsection{AI Use}

The interviews were especially successful in eliciting a variety of real-world use cases for which the student interviewees employ AI technologies. 11 of 13 students mentioned using AI large language models (LLMs) like OpenAI's ChatGPT or Google's Gemini. Explicit mention of the use of AI assistants for coding (e.g. Microsoft CoPilot) or writing (e.g. Grammarly) was less common (4 of 13 interviews). Nevertheless, programming was repeatedly highlighted as a primary use case for AI, even if the interviewee only used prompt-based AI tools like ChatGPT. Nick explained his workflow of using ChatGPT to structure his approach to programming:

\begin{quote}
    ``So let's say there's an assignment with a bunch of requirements. I like to ask [ChatGPT]: What should I start with?...What's a good workflow for that assignment specifically?''
\end{quote}

\noindent The use of LLMs for structuring thoughts or work was also emphasized, with Brett describing ChatGPT as ``a good brainstorming tool,'' and Aidan and Emmanuel using LLMs to organize or refine writing. Likewise, Kenny stressed that he used LLMs by asking ``one question to give me some ideas to get started and then from there I just do my own work,'' cautioning that ``if you've become too reliant on [AI], then you're not going to learn anything yourself.'' Josh emphasized that using Microsoft CoPilot in his programming ``makes everything a lot more efficient,'' since he can use CoPilot to debug his code: ``if I make changes on my own, I can send it into CoPilot [to] debug in case there's any logical problems.''

Many interviewees talked openly about using AI on course assignments, with 8 of 13—both undergraduate and graduate students—noting that the use of AI tools has begun to be encouraged in some of their courses, at least in specific scenarios. Zane mentioned being allowed to use AI to generate outlines for essays in a technical writing course. Similarly, Melissa noted that in one of her courses, the instructor has encouraged the use of ChatGPT to answer questions about course concepts ``because it's really good at breaking down complex topics'':

\begin{quote}
    ``If it’s this sort of multi-layered question, then ChatGPT is really useful for breaking it down: how to solve something and explaining why you need to do certain things.''
\end{quote}

\noindent While Tamara agreed that ``tools such as ChatGPT, Gemini, Claude—they're very helpful for understanding basic concepts,'' she expressed concern that ``a lot of students [are] trying to just get their assignments done through these tools.'' Tamara's skepticism about the use of AI in coursework underscores that views about the integration of these tools into the classroom are still heterogeneous, even among CS students.

Another common use case for LLMs is as an enhanced search engine. For Francisco, ChatGPT is ``just an easier Google sometimes.'' Lorenzo treats an LLM ``more as a fancy Google search engine, where I have this idea in mind, but I'm not entirely sure how to do it in this particular [programming language].'' Zane noted that AI has ``completely replaced Google for me...I go strictly to AI now,'' because he perceives that ``Google and other search engines have gotten worse. At this point it feels like they're cluttered with AI nonsense.''

Regardless of the particular use case, most interviewees viewed AI technologies as beneficial for improving productivity or streamlining workflows, both in a university context and in day-to-day life. The word ``tool'' was used often (by 6 of 13 interviewees) to describe AI, with Corey concluding that ``Ultimately, it's like a helping hand—it's a tool which you will use to save time,'' and Lorenzo describing AI as ``a medium to achieve what you want.'' In sum, then, these responses highlight a steady shift towards the adoption of AI tools among computing students, but do not necessarily imply a wholesale change in the \textit{nature} of student work. AI may improve efficiency, but it is no `one-stop'—much less  `one-click'—solution for completing assignments.

\subsubsection{Ethical Impacts of AI}

While not asked specifically to list specific ethical impacts, Interview Question 3—which asks interviewees how often they ``think about the ethical side of the AI technologies [they] use?''—nevertheless elicited thoughts on specific challenges related to AI, highlighting both the myriad social implications of the technology, and that students think about `ethics of AI' in diverse ways. One common thread, however, was the impact of AI and automation on the workforce, which was brought up by 6 of 13 interviewees (often before IQ3 was asked). Brett and Spencer both noted the potential detrimental effects of pervasive AI on jobs, with Spencer suggesting that ``the nature of jobs is going to change. There's going to be more creative jobs.'' This sentiment was echoed by Emmanuel and Lorenzo, who drew parallels with other technological innovations throughout history. Lorenzo explained:

\begin{quote}
    ``I can understand [people's concern about job loss], but I feel like it's similar to a lot of technologies throughout history, where, even though [technology] replaces some jobs, it's also creating new ones. And so the question remains as to whether [AI will] be able to create enough jobs to replace the ones it got rid of.''
\end{quote}

\noindent This context seems to be a reason why the views of many interviewees on AI remain largely positive despite a recognition of the impacts of automation on the workforce. Yet concern was not completely absent. For Josh, who was looking for a full-time computing job after graduating, his concerns about the impact of AI on the CS job market in particular had tangible near-term connections.

Other ethical impacts of AI mentioned by interviewees included the risk of becoming over-reliant on AI; AI hallucination and mis- or disinformation; issues over intellectual property and content ownership; and the impact of AI model training and deployment on the environment. Melissa, concerned about the latter, explained, “I try not to use [AI] often, just because it does have a pretty significant impact energy-wise, and I don’t love contributing to it.'' 

Yet the continued use of AI tools in spite of the ethical concerns raised by the interviewees was a common theme. For some students, these concerns influence how they engage with AI. For example, Aidan noted that he tends to only use AI for ``simple things'' and ``won't ask it to help me [with] a homework solution.'' Nevertheless, Melissa noted that she still uses AI occasionally ``because I think it is a really useful tool.'' Corey, who uses \textit{ScholarGPT}\footnote{\href{https://chatgpt.com/g/g-kZ0eYXlJe-scholar-gpt}{ScholarGPT} by Sider.ai} in his research, mentioned that he always verifies its output using ``the official Google Scholar'' results.

For other students, however, ethical implications play even less of a role in whether and how they use AI. For Zane, the productivity and efficiency benefits of AI heavily outweigh any drawbacks. He noted, ``I would use AI regardless. I'm not going to choose some other source over AI due to ethical implications.'' Likewise, Lorenzo concluded that ``for the most part, I'm okay with just, you know, trusting [companies like OpenAI, Google, and Microsoft]'' when it comes to AI. Spencer noted, ``I think about ethics broadly. But I wouldn't say I think about it when I'm using day-to-day technologies'' like ChatGPT or Grammarly. This tension was best summarized by Brett, who stated:

\begin{quote}
    ``I think as a computer science major, [I think about the ethics of AI] probably more often than the average person, but even then not that much...Just using [AI] on a day-to-day basis, it's not easy to see the ethical issues, even though I know they're there. It's hard to think about it when you're just interacting with the [user interface].''
\end{quote}

\noindent Asked whether ethical concerns impact his use of AI tools, Brett responded simply, ``Not really, no.'' 

\subsubsection{AI Ethics in CS Courses}

At our university, there is a single undergraduate computing ethics course, \textit{Professionalism in Computing} (hereafter, ``CS ethics course''), which is required for the CS major. The graduate equivalent of the class fulfills a graduate course requirement. As such, a majority of interviewees had taken or were taking the CS ethics course, and many drew from their experiences in that course in discussing ethical implications of AI. In particular, interviewees recalled lectures on intellectual property and generative AI, and remembered discussing data privacy and bias issues in the context of machine learning models. There was less agreement among interviewees when it came to if \textit{AI ethics} is integrated into the curriculum apart from the CS ethics course. Tamara noted that discussions around issues of data privacy, model poisoning, and bias and fairness in other classes ``are rarer but I would say that they're there,'' and Brett mentioned that ethical implications were brought up in \textit{Introduction to AI} and a senior seminar class. By contrast, Kenny and Melissa noted that ethics didn't feature in any of their other courses.

For students who had taken the CS ethics course, many appreciated the content on AI being included in the course curriculum, as it gave them a new perspective on the technologies they use. Nick, for example, reported that ``I don't trust AI as much,'' and ``I'm more cautious'' after learning about AI in the course. While Francisco didn't mind discussing the impacts of AI in the course because he had ``never really thought about it too much,'' he didn't think the course had the full intended effect, adding, ``I don't think that many students are really worried about [AI] yet...especially with everything else they got going [on], professionalism classes are probably the least of their worries.'' Francisco also pointed to the irony of having ``a whole unit on [AI]—not to use it that much and to be weary of it,'' only to have written assignments in the course being graded using an online AI feedback tool. Separately, Kenny was discouraged by the fact that, when talking about AI ethics, ``there's always open ended questions that you can never actually get a definitive answer to.''

Many interviewees also had opinions on how ethics was incorporated into the computing curriculum. Spencer felt there wasn't enough discussion of ethics throughout the whole curriculum, adding that ``I feel like ethics is thrown to the end of the course.'' This view was not shared by all interviewees, with Zane opining:

\begin{quote}
    I think there's enough [ethics in the curriculum]. I'm going to have to take another ethics course for my graduate program. And now I have to take two ethics courses for my whole studies. I feel like it's a little too much ethics. I feel like I could just pursue that on my own without having to take a class for it.''
\end{quote}

\noindent Other students took a more nuanced approach, recognizing the challenges of integrating ethics into a fully-packed technical computing curriculum. Brett and Emmanuel both acknowledged the limited available time in courses and considered whether adding ethics in technical courses would be an effective use of course time. Reflecting on the use of a standalone ethics course in our CS program, Lorenzo concluded:

\begin{quote}
    ``While ethics is important to touch on, it's not the focus of [a technical] class. So I think it's actually good that [university] is requiring a class that completely focuses on ethics.''
\end{quote}

\subsubsection{Government Regulation of AI}

Knowledge of and opinions on AI regulatory efforts by the government were even more sparse than strong opinions on the integration of AI ethics in the CS curriculum. Some students—like Melissa and Nick—recalled relevant conversations on government regulation in the CS ethics course, whereas others—like Corey and Zane— were unaware of current government efforts, with Zane concluding, ``I guess I haven't looked very much into it, so I can't really say it's important to me.'' Josh noted that there are not ``in-depth talks'' about AI regulation in CS courses, adding, ``there is definitely room to improve on educating people...since [ethics and regulation are] barely talked about.'' When students were asked if they would be able to have a conversation with a peer about AI regulation, Kenny responded that he couldn't have a conversation for more than a few minutes, ``because I feel like my knowledge doesn't stretch that far.''

Responses were also mixed with respect to opinions on how the government should regulate AI. Melissa mentioned her concern for ``how prevalent AI disinformation is becoming and already is,'' adding that she would like to see government regulation for AI generated content. She concludes that AI is ``only going to become more and more used as weapons of propaganda, so it's important to crack down on that now.'' Both Brett and Spencer expressed concern that the government lacks ``the necessary expertise'' to regulate AI technologies effectively. On the other hand, Spencer and Emmanuel disagreed that additional regulation of AI is needed. Spencer opined, ``I don't think we need regulation of AI...what I believe we need is regulation of AI organizations,'' while Emmanuel placed emphasis on using existing laws (e.g. copyright law) to steer AI deployment. A third set of students felt discouraged by the lack of effective regulatory measures already in place. This opinion was most clearly expressed by Francisco, who stated, ``There's always been rules against selling data—or laws, I think at least—and companies have still been doing it and getting caught. So regulating [AI]—I don't know how much it would do.''

\subsubsection{AI Career Paths}

We ended each interview by asking the interviewee if they would be interested in pursuing AI policy as a possible career path. In response, only two students (15\%) expressed interest, three students (23\%) expressed an openness to considering it, and the remainder (62\%) indicated that AI policy was not a career for them. Among students in the latter group, many cited a lack of interest in politics and lawmaking as a reason. Brett's response is fairly illustrative here. Although he noted that, ``Personally I think [AI ethics and policy] is one of the more interesting aspects of the technological frontier,'' when asked if he would consider AI policy as a potential career, he replied:

\begin{quote}
    ``Probably not, just because I don't enjoy actually being a part of the politics. I like knowing what's going on—I think it's interesting what happens with it. But in terms of actually making the policy and especially trying to persuade people to agree with you on certain things, it's just too frustrating for me.''
\end{quote}

\noindent Generally, while there was a widespread recognition of the need for AI policy and for policymakers knowledgeable about the technical details of AI technologies, this did not impact students' interest in these potential career trajectories. In other words, interviewees imply that the policymaking process needs more technologists, but they don't imagine themselves as being those technologists.

\section{Discussion}\label{discussion}

Using our interview findings to support broader trends in the survey data, we can begin to identify common themes in CS student attitudes and competencies with respect to AI. Below, we elaborate on three primary themes, placing our findings in the larger context of CS education and the proliferation of AI systems in everyday settings. The first theme relates to \textbf{RQ1}, while the latter two each contribute to \textbf{RQ2} and \textbf{RQ3}.

\subsection{AI as a Tool}

We find that CS students are avid and regular users of AI, with nearly two-thirds of respondents indicating they use AI tools at least a few times per week in their university studies, and nearly three-quarters using AI at least a few times per week in day-to-day life. Compared to the 43\% of young Americans in the Pew \textit{American Trends Panel} survey \cite{mcclain_americans_2024} who indicated having used ChatGPT at least once before, and the less than 25\% of STEM students in the Kim et al. survey \cite{kim_examining_2025} who reported using generative AI tools at least sevel times per week, our finding of remarkably high AI use among CS students suggests (a combination of) two conclusions. First and likely most significant, our results suggest that CS university students are at the vanguard of adopting and adapting to new consumer AI technologies. As such, there is perhaps even more of an important distinction to be made—with respect to AI—between CS students and non-CS students than the difference Kim et al. report in usage and attitudes between STEM and non-STEM students. Second, we note that the Kim et al. survey was fielded in fall 2023, while our survey was fielded a year later. As such, we leave open the possibility that AI usage among the overall student population has increased from 2023 to 2024, potentially leading us to slightly overstate the difference between CS and non-CS student usage. This latter hypothesis is supported by the rapid increase in ChatGPT usage among respondents aged 18-29 between the 2023 and 2024 iterations of the Pew survey \cite{mcclain_americans_2024}.

The widespread utilization of AI tools among our survey population in academic contexts is less surprising given that a wide array of CS courses at our university are beginning to allow or encourage the use of AI tools in classroom settings. Three in five students we interviewed noted that they have taken or are taking a class where this is the case, providing examples of encouraged use cases including using an LLM to clarify complex course topics, using a built-in AI coding assistant on programming assignments, or having an AI tool review and improve student writing. Regarding the latter, we find support for the conclusion from Levine et al. \cite{levine_how_2024} that AI writing tools are predominantly used to improve—not replace—student writing. The adoption of AI in university courses, even if primarily only in computing courses, signals a shift in educators' attitudes regarding the benefits and potential drawbacks of AI. A comprehensive review of how AI is used in the classroom (and what policies towards AI use instructors adopt) appears to us as a promising direction for future research.

More fundamentally, however, our interviews suggest that a widespread adoption of AI technologies as part of students' `productivity tool kits' is taking place even without instructor encouragement. Many interviewees used the word ``tool'' to describe the AI systems they use, and the use of LLMs as a search engine emerged as a common functionality. For some, LLMs have begun to entirely replace conventional search engines. One student remarked that Google, in trying to embed AI-generated content into its search results, has become ``worse'' to use, ironically describing Google's AI-generated content as ``nonsense''. While student use of AI tools isn't without caution—some interviewees detailed the steps they take to verify the outputs they get from AI models—our findings do suggest a substantial change in \textit{how} students approach their coursework. This has implications not only within academia—as educators grapple with what should constitute an `appropriate amount' of AI—but also for industry. On the one hand, these students will be well-prepared and accustomed to using effective AI tools in their workflow. At the same time, however, these students may expect a workplace saturated with AI tools, thereby developing a conception of the workplace which may not conform to current or future reality.  

\subsection{AI and Its Social Impacts}

The outlook of CS students towards AI is exceedingly positive, with students viewing current AI technologies as providing more benefits than drawbacks. The percentage of respondents agreeing to this statement increases from 73\% to 79\% when referencing \textit{future} AI technology. In other words, respondents believe that the technical capabilities of AI will continue to improve over time, while harms subside. This is evidenced by the survey responses for SQ2.9 and 2.10, which asked students to select the two potential impacts of AI they worry about the most. The two most common concerns, deepfake content and data privacy, saw 20\% and 25\% decreases, respectively, in selection when talking about \textit{future} worries, whereas bias, discrimination and/or stereotyping saw a 42\% decrease. Conversely, concern about loss of jobs due to AI-driven automation, the impact of AI on human emotions and behavior, and increasing wealth inequality all saw increases between the `current' and `future' frames of reference. We posit that these trends reflect a distinction between \textit{technical} AI problems, which students may anticipate being mitigated or solved through technical advancements in AI content detection and data privacy methods, and \textit{social} impacts of AI, which students anticipate will be more relevant in the future as AI adoption increases. 

In one sense, it is reassuring that students acknowledge the potential for AI to have large-scale ramifications on our social fabric. Indeed, the top ethical concern expressed by students in our interviews—and by students in the Ghotbi and Ho study \cite{ghotbi_moral_2021}—was the impact of AI automation on jobs. The risks of over-reliance on AI, hallucination, issues of intellectual property, and the impact of AI models on the environment were also discussed in our interviews. We also emphasize that exposure to those topics in CS ethics courses seems to heighten students' ethical awareness when using AI technologies. 

On the other hand, we worry that CS students view the `technical' problems of deepfakes, misinformation, bias, and privacy as issues that will \textit{inevitably} be solved (by someone else), and not fundamental challenges to the domain of AI and ML which require conscious efforts and contributions from \textit{all} practitioners. We therefore find it imperative that AI (and CS) curricula not only highlight ethical challenges of AI—including questions about intellectual property and content ownership related to generative AI training and output \cite{custers_generative_2022, quintais_generative_2025}, latent inequities in training data that lead to bias and discrimination \cite{buolamwini_gender_2018}, the carbon footprint of training and deploying large language models \cite{luccioni_2023}, etc.—but also to talk about and integrate into assignments proposed technical solutions to these challenges. Students must develop familiarity with translating ethical principles to practices \cite{kim_exploring_2023, schiff_principles_2020} beginning not at the workplace but at the university.     

Regarding student competencies with respect to AI, survey responses indicate that students have confidence in their ability to explain how most AI technologies work (67\% agree or strongly agree with SQ1.2) and how AI systems can make biased or harmful decisions (88\% agree or strongly agree with SQ2.5). These students are less confident in their ability to \textit{mitigate} potential harms (66\% agree or strongly agree with SQ2.8), and less than half of respondents feel prepared to engage in conversations about AI policy and regulation (46\% agree or strongly agree with SQ3.8). We argue that this is a direct consequence of a CS curriculum that places the most emphasis on technical understanding, less emphasis on engagement with ethical impacts \cite{weichert_i_2025}, and barely any emphasis on shaping or responding to discussions concerning the regulation and oversight of technology \cite{weichert_evolution_2025}. We question whether this prioritization fully aligns with what society expects of its computer science graduates, or whether changes to the structure of the curriculum are needed to better prepare students for the complicated ethical and policy realities of the future.

\subsection{The Politics of AI}

The survey data and interview responses reveal that, while most students believe more government regulation of AI is necessary (59\% agreement in SQ3.1) and express desires for regulation to protect both users and society (SQ3.3 and 3.4), students differ on exactly what AI technologies should be regulated and how. Views expressed by interviewees ranged from desires for specific AI model requirements (e.g. AI content watermarks) to regulation of AI organizations (as opposed to AI models) to using existing laws to govern AI technology. A majority of students, however, either indicated a lack of knowledge on the subject, or a lack of confidence that regulation would be effective at mitigating AI harms.

One of the primary motivations for this study was to determine the proportion of CS students who are interested in or would be open to a career involving AI-related policy work. The survey and interviews suggest that this figure is around 30-40\%. In this respect, these findings support our conclusion from our initial study \cite{weichert_computer_2024} that although AI policy is of interest to a not insignificant number of CS students, the topic is insufficiently incorporated into the CS curriculum, both in computing ethics courses and in technical AI/ML courses. We also emphasize the conclusion of Kim and Katz \cite{kim_engineering_2025} that introducing students to career opportunities beyond the private sector is likely to ``help broaden participation'' in fields like technology policy, ``especially by attracting students with diverse backgrounds and motivations,'' That some students express strong opinions on AI policy despite the under-representation of AI regulation as a course topic would suggest that students are forming these opinions outside of the classroom in their role as AI \textit{users}. This is a further reason why our `dual conceptualization' of CS students as \textit{AI learners} and \textit{AI consumers} is useful in understanding the CS student population in particular.

More broadly, though, our analysis of interview responses highlights two key reasons why students aren't more interested in AI policy. First, multiple interviewees described their disillusionment with policymaking and the political process stemming from a lack of positive impacts of government policy on their own lives or the belief that—as expressed by one student—regulation `doesn't do much' in the face of multi-billion dollar technology companies, legal loopholes, and deceptive practices. These opinions manifest in a disinterest in getting involved with a political system that young people may increasingly feel has failed them \cite{wattenberg_is_2024}. While interviewees recognized that policymakers do not currently have the expertise necessary to be effective at regulating AI and that the incorporation of AI experts is needed in the policymaking process, they are reluctant to assume these responsibilities themselves. This attitude is best summarized by one interview response: ``Not for me, probably for someone else.'' 

Second, the interview conversations about policy and regulation hinted at some students' discomfort when faced with normative, open-ended questions. Whereas the field of computing conventionally demands precise, succinct, and concrete answers, ethical challenges confront students with what one interviewee termed ``non-deterministic scenarios,'' requiring contextualized, relativist, or `fuzzy' approaches. There is potential, then, for students to become frustrated when they attempt to apply traditional computing `ways of knowing' to ethical ways of knowing grounded in philosophical theory. Here we echo the call by Raji et al. \cite{raji_you_2021} for AI ethics to move away from ``exclusionary'' pedagogical practices that reinforce disciplinary silos and promote ``the engineer’s natural inclination towards
seeing themselves as a solitary saviour, to the detriment of the
quality of the solution and in spite of the need for other disciplinary perspectives.'' In other words, CS ethics courses should build student comfort with \textit{not knowing} the `right answer' and should reinforce interdisciplinary approaches to designing and implementing equitable computing systems.

There is disagreement among students, as there is in the literature, about where and how ethics content should be incorporated into the CS curriculum. Yet only one student we interviewed thought that there should be \textit{less} of a focus on AI ethics and policy instead of more. In sum, we believe this paper provides yet another justification for an increased curricular focus on the social impacts of AI, how these impacts can be addressed, and how AI can be responsibly regulated. This content is not only relevant to the third of students who may be interested in AI policy-related career paths, but equally as beneficial for more `technically-minded' students who may find that their future AI job requires them to navigate the AI policy landscape and incorporate ethical principles or regulations into the design and construction of AI models. 

\subsection{Limitations}

While we do not validate our survey instrument as part of this study, we use the survey data as a guide for understanding general trends among the CS student population. In particular, we compare response distributions between multiple questions in the survey. We also augment our quantitative survey data with in-depth student perspectives elicited through our follow-up interviews. By evaluating common themes across the 13 interviews, we can construct a more comprehensive understanding of how student attitudes towards AI are shaped through their experiences in and outside of the classroom. Likewise, interview responses also allow us to identify cases where the survey responses do not capture the `full picture,' or where additional context is needed.

\section{Conclusion}

This study presents an overview of attitudes and competencies related to AI, AI ethics, and AI regulation among undergraduate and graduate CS students at a large public university in the United States. We combine the findings of an online survey of 117 students with an analysis of 13 semi-structured interviews to identify common themes across responses relating to the use of AI tools, attitudes towards the potential ethical challenges of AI, and interest in career paths related to AI policy. We find that CS students have a largely positive outlook on AI, influenced in large part by their widespread adoption of AI tools ranging from ChatGPT to Grammarly to social media recommendation algorithms. Nevertheless, students are cognizant of the impacts and potential harms of these technologies, with a plurality emphasizing their concern for the loss of jobs in the future due to AI-driven automation. Crucially, we substantiate the finding from our previous survey that around a third of CS students may be open to pursuing a career involving AI policy work. We conclude by addressing the implications of these findings for the CS curriculum, particularly as it relates to AI and computing ethics courses. While  this study focuses on academia, our findings and conclusions should nevertheless be of interest to leaders in industry and politics, especially with respect to student use of AI tools in their workflows and student attitudes towards government regulation of technology.

\backmatter 

\bigskip

\begin{appendices}

\clearpage

\section{Survey Questions}\label{secA1}

\begin{table}[!htb]
    \centering
    \centerline{

    \begin{tabular}{c|p{14cm}|c}
        
        \toprule
        \textbf{SQ} & \multicolumn{1}{c|}{\textbf{Question}} & \textbf{Type} \\
        
        \midrule
        \multicolumn{3}{c}{\textbf{\textit{General Attitudes Towards AI}}} \\
        \midrule

        1.1 & I am familiar with modern artificial intelligence (AI) technologies like neural networks, large language models (LLMs), and autonomous vehicles (“self-driving cars”). & \multirow{11}{*}{Likert} \\

        1.2 & I can explain to someone who is not familiar with AI, in general terms, how most AI technologies work. & \\

        1.3 & I believe that current AI tools provide more benefits than drawbacks. & \\

        1.4 & I believe that future AI tools will provide more benefits than drawbacks. & \\

        1.5 & Much of society will benefit from a future full of AI.\footnotemark & \\

        1.6 & AI tools are helpful to me in my day-to-day life. & \\

        1.7 & AI tools are helpful in my university studies. & \\

        1.8 & Using AI tools increases my productivity. & \\

        \midrule

        1.9 & How often do you use AI tools (e.g. ChatGPT, Google Gemini, Tesla Autopilot, etc.) in your day-to-day life? & \multirow{4}{*}{Multiple Choice} \\

        1.10 & How often do you use AI tools in your university studies?  & \\

        1.11 & What, if any, AI technologies do you use? & \\

        \midrule
        \multicolumn{3}{c}{\textbf{\textit{AI Ethics}}} \\
        \midrule

        2.1 & In general, I think existing AI tools are ethical. & \multirow{12}{*}{Likert} \\

        2.2 & I believe that most developers of AI tools design their AI systems with ethics in mind. & \\

        2.3 & I worry about the ethical impact of current AI technology. & \\

        2.4 & I worry about the ethical impact of future AI technology. & \\

        2.5 & I can explain to someone not familiar with AI how AI can make biased or harmful decisions. & \\

        2.6 & My artificial intelligence and/or machine learning courses at university integrate discussion about the ethics of AI into the curriculum. & \\

        2.7 & My computer science education focuses sufficient attention on AI ethics. & \\

        2.8 & My courses at university have prepared me to discuss and mitigate potential harms that AI can cause. & \\

        \midrule

        2.9 & Which of the following potential ethical impacts of AI technology do you most worry about now?  (select up to 2) & \multirow{4}{*}{Select Multiple} \\

        2.10 & Which of the following potential ethical impacts of AI technology do you think will be most important in the future? (select up to 2) & \\

        \midrule
        \multicolumn{3}{c}{\textbf{\textit{AI Regulation and Policy}}} \\
        \midrule

        3.1 & I believe AI technologies are currently adequately regulated by the government. & \multirow{14}{*}{Likert} \\

        3.2 & The U.S. government is doing a good job of balancing promoting innovation and protecting users with respect to AI technology. & \\

        3.3 & The U.S. government should do more to protect users from the potential harms of AI technology. & \\

        3.4 & The U.S. government should do more to protect society from the potential harms of AI technology. & \\

        3.5 & I tend to follow news about government regulation of technology and/or AI. & \\

        3.6 & I am interested in AI policy and regulation as a potential career path. & \\

        3.7 & We talk about the regulation of AI by the government in my artificial intelligence and/or machine learning courses at university. & \\

        3.8 & My courses at university are adequately preparing me to engage in conversations about AI policy and regulation. & \\

        \bottomrule
    \end{tabular}
    
    }
    \caption{Survey questions organized by section.}
    \label{tab:survey-questions}
\end{table}

\footnotetext{Adapted from Schepman and Rodway \cite{schepman_general_2023}.}

\clearpage

\section{Interview Prompts}\label{secA2}

\begin{table}[!htb]
    \centering
    \begin{tabular}{c|p{10cm}}
         
         \toprule
         \textbf{Question} & \multicolumn{1}{c}{\textbf{Prompt}} \\
         \midrule

         IQ1 & Could you describe your general feelings towards artificial intelligence technologies? Do you think that AI is currently a net positive or net negative to society? \\

         IQ2 & Do you use any AI tools in your day-to-day life as a student or outside of the university? If so, what tools do you use and what does your ‘AI workflow’ look like? \\

         IQ3 & How often would you say you think about the ethical side of the AI technologies you use? Is it something that impacts your decisions about whether to use AI or not? \\

         IQ4 & How would you describe how the ethics of AI is approached or talked about, if at all, in your computer science courses? What are your opinions on this approach? Do you think we talk enough about AI ethics in these courses? \\

         IQ5 & How much do you think about what the government is doing to regulate AI? Is government regulation of AI something that’s important to you? Do you think there is currently adequate regulation of AI? \\

         IQ6 & Is AI policy or regulation talked about in your courses? Is it something that you feel able to have a conversation about? Is AI policy an area you’d be interested in pursuing as a possible career path? \\

         \bottomrule
    \end{tabular}
    \caption{Follow-up interview prompts.}
    \label{tab:interview-questions}
\end{table}

\clearpage

\section{Likert-Scale Survey Responses}\label{secA3}

\begin{figure}[!htb]
    \centering
    \includegraphics[width=1\linewidth]{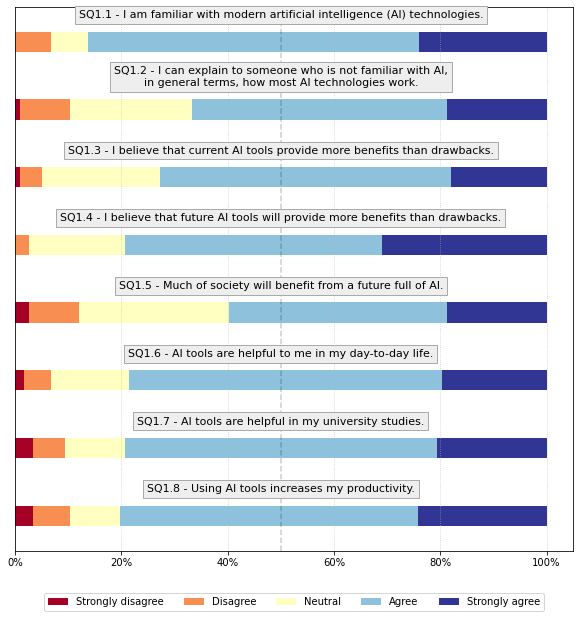}
    \caption{Survey responses to Likert-scale questions about \textbf{general attitudes} towards AI.}
    \label{fig:general-likert}
\end{figure}

\begin{figure}[]
    \centering
    \includegraphics[width=1\linewidth]{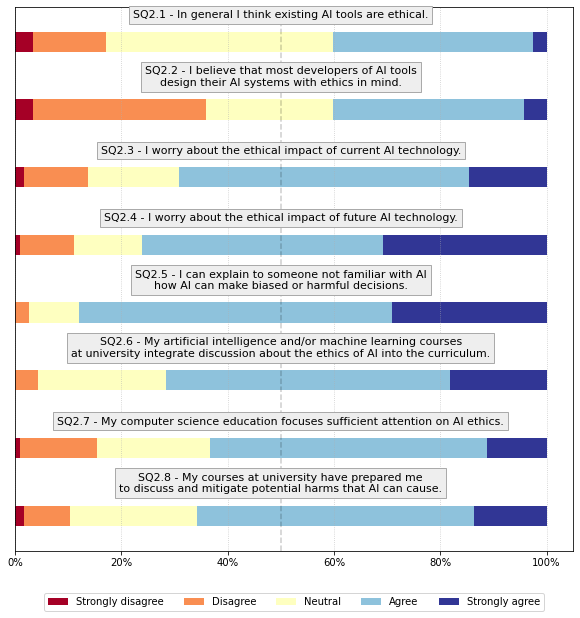}
    \caption{Survey responses to Likert-scale questions about the \textbf{ethics of AI}.}
    \label{fig:ethics-likert}
\end{figure}

\begin{figure}[]
    \centering
    \includegraphics[width=1\linewidth]{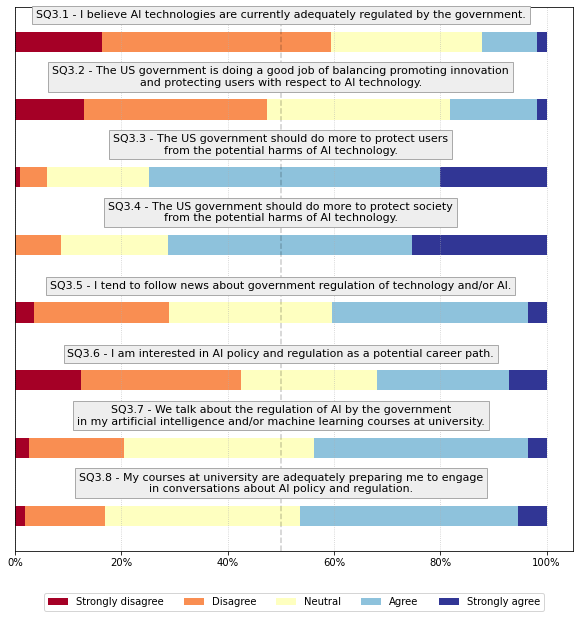}
    \caption{Survey responses to Likert-scale questions about \textbf{AI policy and regulation}.}
    \label{fig:policy-likert}
\end{figure}




\clearpage

\end{appendices}


\bibliography{references}

\end{document}